\documentclass[fleqn,twoside]{article}
\usepackage{espcrc2}
\usepackage{a4wide}

\usepackage{graphicx}
\usepackage[figuresright]{rotating}

\newcommand{\AmS}{{\protect\the\textfont2
  A\kern-.1667em\lower.5ex\hbox{M}\kern-.125emS}}

\title{Saturation effects in $pp$ scattering
in the impact-parameter representation\thanks{presented by O.V.S.
at ``Diffraction 2004'', 
International Workshop on Diffraction in High-Energy Physics, 
   Cala Gonone, Sardinia, 
   September 18-23, 2004.} }

\author{
 O.V. Selyugin$^{1,}$\thanks{on leave from the Bogoliubov
 Laboratory of Theoretical Physics, JINR, 141980, Dubna, Moscow Region,
 Russia, e-mail:selugin@qcd.theo.phys.ulg.ac.be.}
and J.R. Cudell\thanks{Institut de Physique, B\^at. B5a,
Universit\'e de Li\`ege, Sart Tilman, B4000
  Li\`ege, Belgium, e-mail: J.R.Cudell@ulg.ac.be }
}
\begin{document}

\begin{abstract}
The impact of unitarity is considered in different approaches 
to saturation in impact-parameter space.
The energy and momentum-transfer dependence of the total and differential
cross sections and of the ratio of the real to imaginary parts of the scattering
amplitude are obtained in a model that includes soft and hard pomeron
contributions, coupled to hadrons via the electromagnetic form factor.
It is shown that the hard pomeron may significantly contribute
to soft physics at the LHC. A similar conclusion can also be reached
in the framework of non-linear approaches to unitarisation of the BFKL pomeron.
\vspace{1pc}
\end{abstract}

\maketitle

\section{Introduction}
One expects that non-linear effects will enter the BFKL equation
in the non-perturbative infrared region, {\it i.e.}  
at large impact parameters. This is a different regime from
that connected with the Black Disk Limit (BDL), in which saturation occurs
at small impact parameters first.

The common viewpoint is that saturation will lead
to a decrease of the growth of $\sigma_{tot}$.
But the estimates of the energy after which saturation will be important
vary a lot between different models.

 Unitarity of the scattering matrix is connected with
the properties of the scattering amplitude in the impact parameter
 representation, which is equivalent at high energy
to a decomposition in partial waves.
The scattering amplitude 
can then saturate the unitarity bound for impact parameters
 $b<b_i$.
To satisfy the unitarity condition, there are different
  models. Two of them are based on the solution of the unitarity equation
$SS^\dagger=1$.
First of all, in the $U-$matrix approach~\cite{trosh},
one obtains a ratio
 $         {\sigma_{el}}/{\sigma_{tot}} \ \rightarrow \ 1, $
as $s\rightarrow \infty$.
The second possible solution of the unitarity condition, which is the usual one,
corresponds to the eikonal representation
\begin{eqnarray}
  T(s,t) = i \int_{0}^{\infty}  b db J_{0}(b \Delta)
  \left[1 -  \exp(-\chi(s,b))\right]
\end{eqnarray}
with $t=\Delta^2$.
  If one takes the eikonal phase in a factorised form
$ \chi(s,b) = h(s) \ f(b)$,
one  usually supposes that, despite the fact that the energy dependence of
$h(s)$ can be a power
$ h(s) \sim s^{\Delta}$,
the total cross section will satisfy the Froissart bound
$            \sigma_{tot} \leq \ C \ \log^2 (s)$.

We find in fact that
the energy dependence of the imaginary part of the amplitude
and hence of the total
cross section depends on the form of $f(b)$, {\it i.e.} on
the $s$ and $t$ dependence of the slope of the elastic scattering
amplitude.

\subsection{Eikonal representation and  $\sigma_{tot}$ }

  Let us first take a Gaussian for the eikonal phase 
 $       f(b) \sim \exp(-b^2/R^2)$.
 As was made  first by Landau, let us
  introduce the new variable
$ y \ = \ \exp(-b^2/R^2)$.
 We can then calculate the integral
  exactly  and obtain that
\begin{eqnarray}
  T(s,t) \sim i \ R^2 \ \left[\Gamma(0,s^{\Delta}) \ + \ \Delta \log{s}\right],
\end{eqnarray}
 where
$  \Gamma(a,z) = \int_{z}^{\infty} \ t^{a-1} \  e^{-t} \ dt $
 and, in our case,
$  \Gamma(0,s^{\Delta}) \rightarrow  \ 0$ as $ s  \rightarrow \infty $ .
  If $R^2$ is independent from $s$, we have
$ \sigma_{tot} \ \sim \  \log(s) $,
  whereas if $R^2$ grows not faster then $ \log(s)$, the total cross
section becomes proportional to the Froissart bound
 $ \sigma_{tot} \  \sim  \ \log^2(s)$.

  However, let us now take a polynomial form
 $ f(b) \ \sim \ s^{\Delta} /{b^4} $.
  Such a form comes, for example, from the BFKL equation.
  In that case, we can introduce the new variable $y= 1/b^4$
  and obtain
\begin{eqnarray}
  T(s,t) &\sim& i \int_{0}^{\infty} \ \frac{ 1}{y \sqrt{y}}
  \ [1\ - \exp( - s^{\Delta} \ y)] \  \nonumber \\
 &=& \ 2 \sqrt{\pi} \ s^{\Delta/2}.
\end{eqnarray}
  So, in this case, the scattering amplitude
eventually violates the Froissart bound!

If we introduce an additional small constant radius $r$ which removes
the singular point $b=0$  in $f(b)$ and take
 $ f(b) \ \sim \ s^{\Delta}/[b^4+r^4] $,
the answer, after some complicated algebra, is
\begin{eqnarray}
 T(s,t=0) &\sim& \frac{1}{4r^2}\left\{\pi s^{\Delta} 
\exp\left[-\frac{s^{\Delta}}{2r^4}\right]\right\}
 \nonumber \\
 &\times& \left[I_{0}\left(\frac{s^{\Delta}}{2r^4}\right) +I_{1}\left(\frac{s^{\Delta}}{2r^4}\right)\right].
\end{eqnarray}
 The asymptotic value of the  Modified Bessel functions is
\begin{eqnarray}
  I_{0,1}\left(s^{\Delta}/(2r^4)\right) \ \sim \ r^2/ \left(\sqrt{\pi} s^{\Delta/2}\right).
\end{eqnarray}
We again obtain for asymptotically high energies
\begin{eqnarray}
  T(s,t) \sim i \ \sqrt{\pi} s^{\Delta/2}/ \left(2 \sqrt{2}\right).
\end{eqnarray}

\begin{figure}[htb]
\includegraphics[width=7.5cm]{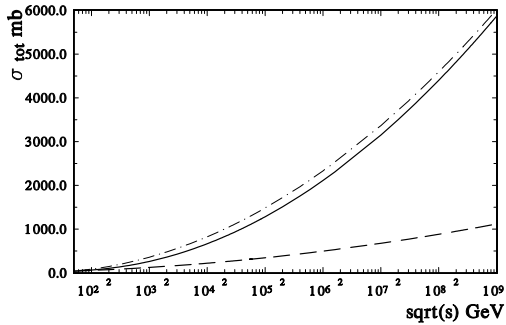}
\caption{
The total cross section of proton-proton scattering
calculated in the eikonal representation (hard line:
    with an exponential form; dashed line: with a Gaussian form)
compared with a $\log^2(s)$ dependence (dash-dotted line).
}
\label{fig:largenenough}
\end{figure}

Finally, let us consider the more complicated (but most interesting) case, 
of an exponential form for 
$ f(b) \sim \exp(-m b) $.
The corresponding amplitude in the $t$ representation is
\begin{eqnarray}
T(s,t) \ \sim \ i \ s^{\Delta} \  _{q}F_{p}[(1,1,1),(2,2,2), -s^{\Delta}].
\end{eqnarray}
with $p=q=3$, and where
the function $ _{q}F_{p}$ is the  hyper-geometric function.

A numerical estimate of this integral
shows that
we obtain for the exponential form of the eikonal
an energy dependence of the scattering amplitude in the $t$-representation
which can be approximated as
 $  T(s,t=0) \ \sim \ i \ a \  \log^2(s/s_0)$,
 with large coefficients $a=4.5$ and $s_0 = 135$ GeV$^2$.

Hence, in this case, the scattering amplitude obeys the
Froissart bound, but with a large scale
$s_0 $ and a large coefficient 
$a$. This leads
to a weak energy dependence at moderate energies followed by a fast growth
at super-high energies.

Of course the eikonal representation, which most of the time
leads to a unitarity answer at every impact parameter, leads to
amplitudes which do not saturate at finite energy:
the eikonal representation for the scattering amplitude
in $b$-space, in the form $1-\exp(-\chi(s,b))$, reaches
the BDL only asymptotically. However, this representation is not the
only possibility, and it may be more useful to consider the effects
of saturation by considering parametrisations
in $s$ and $t$, transforming them to impact parameter space, and
imposing directly the BDL as an upper bound on the amplitude in $s$ and $b$.

\subsection{Non-linear effects and $\sigma_{tot}$ }
A different approach to saturation 
is found in the studies of the non-linear saturation processes, which have
been considered in a perturbative QCD context \cite{grib,mcler}.
Such processes lead to an infinite set of coupled
evolution equations in energy for the correlation functions of
multiple Wilson lines \cite{balitsky}.
In the approximation where the correlation functions for more than
two Wilson lines factorise, the problem reduces to the non-linear
Balitsky-Kovchegov (BK) equation \cite{balitsky,kovchegov}.

It is unclear how to extend these results to the non-perturbative region,
but one will probably obtain a similar equation. In fact we found simple
differential equations that reproduce either the U-matrix or the eikonal
representation.
We can first consider saturation equations of the general form
\renewcommand{\S}{{\cal S}}
\begin{equation}
 \partial N(\xi,b)/\partial \xi = \S(N)
\end{equation}
with $N$ the true (saturated) imaginary part of the amplitude. We shall
impose that
(a) $N\rightarrow 1$ as $s\rightarrow\infty$,
(b) $\partial N/\partial\xi\rightarrow 0$ as $s\rightarrow\infty$,
(c) $\S(N)$ has a Taylor expansion in $N$, 
with the hard pomeron $N_{bare}=f(b) s^\Delta$ as a first term.
This enables us to fix the integration constant by demanding that the first
term of the expansion in $s^\Delta$ reduces to $N_{bare}$.

Inspired by the BK results, we shall use the evolution variable
$\xi=\log s$. If we want to fulfil condition (c), then we need to
take $\S(N)=\Delta N+O(N^2)$. Conditions (a) and (b) then give
$\S(N)=\Delta(N-N^2)$ as the simplest saturating function.
The resulting equation
\begin{equation}
 \partial N/ \partial \log s = \Delta (N-N^2)
\label{satu1}
\end{equation}
has the solution
\begin{equation}
N= f(b)s^\Delta/ (f(b)s^\Delta + 1)
\label{solU}
\end{equation}
One can in fact go further: eq. (\ref{satu1}) has been written
for the imaginary part of the amplitude.
If we want to generalise  it to a complex amplitude,
so that it reduces to (\ref{satu1}) when the real part vanishes, we must take:
\begin{equation}
 \partial A/ \partial \log s = \Delta (A+iA^2)
\label{satu1c}
\end{equation}
The solution of this is exactly the form 
obtained in the U-matrix formalism, for $\Im U(s,b)=s^\Delta f(b)$.

Many other unitarisation schemes are possible, depending
on the function ${\cal S}(N)$.
We shall simply indicate here that the eikonal scheme
can be obtained as
follows:
\begin{equation}
 \partial N/ \partial \log s=\Delta (1-N) (-\log(1-N))
\end{equation}
Other unitarisation equations can be easily obtained via another first-order
equation. The idea here is that the saturation variable is the imaginary
part of the bare amplitude. One can then write
\begin{eqnarray}
 \partial N/\partial N_{bare}&=&{\cal S}'(N)\Rightarrow  \nonumber \\
 \partial N/\partial \log s&=& \left[\partial N_{bare}/\partial \log s\right]\ {\cal S}'(N)
\end{eqnarray}
with $N_{bare}$ the unsaturated amplitude.

This will trivially obey the conditions (a)-(c) above, and saturate at $N=1$.
Choosing ${\cal S}'(N)=1-N$ gives the eikonal solution
whereas ${\cal S}'(N)=(1-N)^2$ leads to the U-matrix representation
(\ref{solU}).
   We have thus shown that the most usual unitarization schemes could be recast into
  differential equations which are reminiscent of saturation equations
\cite{balitsky,kovchegov}.
  Such an approach can be used to build new unitarization schemes
  and may also shed some light on the physical processes
underlying the saturation regime.

\section{Conclusion}
In the presence of the hard Pomeron \cite{mrt},
the saturation effects
can change the behaviour of some features of
the cross sections already at LHC
energies.
Some forms of the eikonal phase
in the factorising eikonal representation can lead to a violation
of the Froissart bound.
Non-linear effects which work in the whole
energy region  supply an acceptable growth of the total cross sections.
Saturation leads to a relative growth
of the contribution of  peripheral interactions.
The most usual unitarization schemes
could be recast into
differential equations which are reminiscent of saturation equations
Such an approach can be used
to build new unitarization schemes
and may also shed some light on the physical processes
underlying the saturation regime.

\end{document}